\documentclass[reprint,superscriptaddress,showpacs,amsmath,amssymb,aps,prl]{revtex4-1}
\usepackage{graphicx}
\usepackage{dcolumn}
\usepackage{bm}
\usepackage{hyperref}
\usepackage{subfigure}
\hypersetup{colorlinks=true, citecolor=blue, urlcolor=blue, linkcolor=blue}
\usepackage{amssymb}
\usepackage{color}
\usepackage{graphicx}
\usepackage{amsmath}
\usepackage{mathrsfs}
\usepackage[ruled,vlined,linesnumbered]{algorithm2e}
\usepackage{times}
\usepackage{subeqnarray}
\usepackage{cases}
\usepackage{bm}
\usepackage{diagbox}
\usepackage{booktabs}
\usepackage[table,xcdraw]{xcolor}
\usepackage{colortbl}
\usepackage{epstopdf}
\definecolor{tabcolor}{rgb}{.105,.410,.113}
\definecolor{tabcolor2}{rgb}{.100,.149,.237}
\begin{document}

\title{Evolution of cooperation in the public goods game with Q-learning}

\author{Guozhong Zheng}
\affiliation{School of Physics and Information Technology, Shaanxi Normal University, Xi'an 710061, P. R. China}
\author{Jiqiang Zhang}
\affiliation{School of Physics, Ningxia University, Yinchuan 750021, P. R. China}
\author{Shengfeng Deng}
\affiliation{School of Physics and Information Technology, Shaanxi Normal University, Xi'an 710061, P. R. China}
\author{Weiran Cai}
\affiliation{School of Computer Science, Soochow University, Suzhou 215006, P. R. China}
\author{Li Chen}
\email[Email address: ]{chenl@snnu.edu.cn}
\affiliation{School of Physics and Information Technology, Shaanxi Normal University, Xi'an 710061, P. R. China}

\begin{abstract}
Recent paradigm shifts from imitation learning to reinforcement learning (RL) is shown to be productive in understanding human behaviors. In the RL paradigm, individuals search for optimal strategies through interaction with the environment to make decisions. This implies that gathering, processing, and utilizing information from their surroundings are crucial. 
However, existing studies typically study pairwise games such as the prisoners' dilemma and employ a self-regarding setup, where individuals play against one opponent based solely on their own strategies, neglecting the environmental information.
In this work, we investigate the evolution of cooperation with the multiplayer game --  the public goods game using the Q-learning algorithm by leveraging the environmental information. Specifically, the decision-making of players is based upon the cooperation information in their neighborhood. Our results show that cooperation is more likely to emerge compared to the case of imitation learning by using Fermi rule. Of particular interest is the observation of an anomalous non-monotonic dependence which is revealed when voluntary participation is further introduced. The analysis of the Q-table explains the mechanisms behind the cooperation evolution. 
Our findings indicate the fundamental role of environment information in the RL paradigm to understand the evolution of cooperation, and human behaviors in general.
\end{abstract}

\date{\today }
\maketitle
\section{1. Introduction}\label{sec:introduction}

Cooperation is crucial for the sustainable development of humans from tackling climate warming~\cite{Manfred2006Stabilizing}, fighting global infectious diseases, and dealing with the melting of glaciers causing the accelerated extinction of species, where the web of lives that sustains us is being worn and torn apart. The solution to these issues depends on human cooperation~\cite{Kollock1998Social}. Deciphering the cooperation mechanism is thus a question of paramount importance, and is crucial for addressing the challenges of these public tragedies~\cite{Garrett1968The}.

To achieve this aim, researchers investigate the emergence of cooperation by adopting the framework of evolutionary game theory~\cite{nowak2004evolutionary,Nowak2006Five}, where the prisoners' dilemma (PD) game~\cite{Robert1981The} serves as a typical model to study the evolution of cooperation behavior in pairwise interactions~\cite{Doebeli2005Models}.
In the PD, there is a potential conflict for the two engaged players between the individual and their collective interests~\cite{rapoport1965prisoner,Robert1981The}. 
Importantly, the public goods game (PGG)~\cite{Hauert2003dilemma,Perc2013Evolutionary,Jun2013Difference} extends the scenario to multilateral interactions, where the number of participants could be arbitrary. In PGG, cooperation contributes to the common goods, but the free-riding behaviors ruin the benefit of the collective and ultimately lead to the tragedy of commons if no countermeasure is engaged~\cite{Dawes1980Social,Urs2001Are,Wang2009Emergence}. Nowadays, the PGG is widely adopted to discuss many issues we are facing, such as resource management, environmental protection, and the provision of public services~\cite{Herbert2003Explaining}.

After endeavors for several decades, various mechanisms for the emergence of cooperation in the PGG have been proposed, including voluntary participation~\cite{Phase2002Szabo,Hauert2002Volunteering,Christoph2002Replicator,Semmann2003Volunteering}, punishment~\cite{Fehr2002Altruistic,Robert2003The,Perc2017Statistical,
Yang2017Phase} and reward~\cite{Karl2001Reward,Szolnoki2010Reward,Fang2019Synergistic}, social diversity~\cite{Santos2008Social}, noise~\cite{Guan2007Effects}, network heterogeneity~\cite{Rong2009Effect}, reputation and reciprocity~\cite{Xia2023Reputation}, etc. These endeavors provide important insights into the mechanism of cooperation emergence. However, most of these findings are mainly based on imitation learning (IL)~\cite{Nowak1992Evolutionary,Carlos2009Evolutionary}, where players imitate strategies of neighbors who are better off in terms of payoffs. In essence, IL can be taken as a simple version of social learning~\cite{Bandura1977social}, where individuals learn from others in their socioeconomic activities, through observation or instruction, which may or may not involve direct experiences. 
As a different paradigm, reinforcement learning (RL)~\cite{Sutton2018reinforcement} provides a completely distinct framework for understanding human behaviors, which has garnered increasing attention in recent years, including deciphering the emergence mystery of cooperation~\cite{Tanabe2012Evolution,Ezaki2016Reinforcement,
Horita2017Reinforcement,Ding2019Q,Fan2022Incorporating,Zhang2020Oscillatory,Wang2022Levy,Wang2023Synergistic,He2022migration,Ding2023Emergence,Geng2022Reinforcement,Yang2024Interaction,Shi2022analysis,Zhang2024emergence,Zhao2024Emergence}, trust~\cite{Zheng2024decoding}, resource allocation~\cite{Andrecut2001q,Zhang2019reinforcement}, and other collective behaviors for humans~\cite{Tomov2021multi,Shi2022analysis}. There players acquire information through interactions with the environment and adjust behaviors based on rewards provided by the environment, aiming to maximize their cumulative rewards.

\begin{figure*}[htbp!]
\centering
\includegraphics[width=0.8\linewidth]{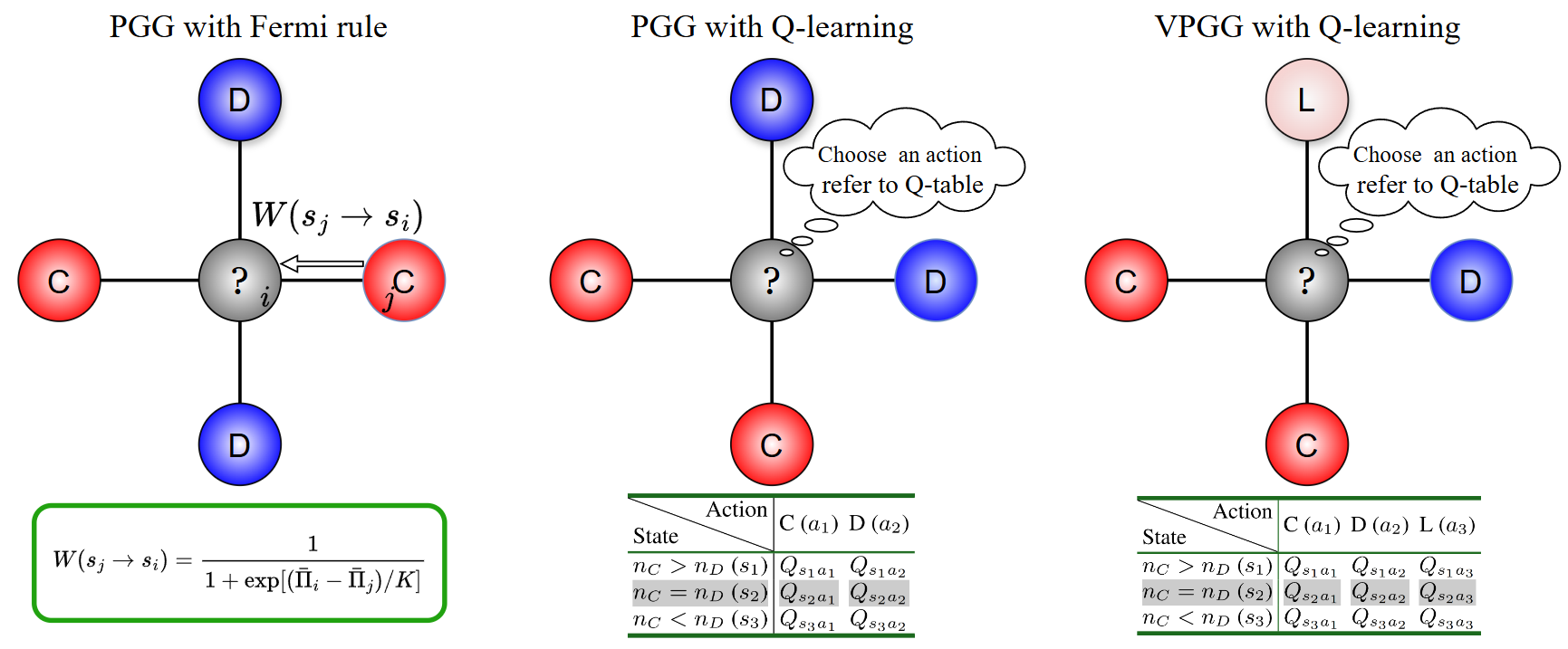}
\caption{{\bf Three different setups for PGG.}
(Left) PGG with Fermi rule: players engage in the PGG and update their strategies using the Fermi updating rule. 
(Middle) PGG with Q-learning algorithm: players engage in the PGG and update their strategies using the Q-learning algorithm. 
(Right) VPGG with Q-learning algorithm: players engage in the VPGG and update their strategies using the Q-learning algorithm.
The grey circle represents the focal individual, and red, blue, and pink represent the cooperators, defectors, and loners in the neighborhood.  
At the bottom of each panel, the shown are the Fermi rule, and the corresponding Q-tables for the other two setups using Q-learning, respectively.
}
\label{fig:model}
\end{figure*}

Within the framework of RL, some new insights into the mechanism of cooperation emergence for the pairwise game have been provided. For example, by adopting the Bush-Mosteller model as a simple RL setup, Masuda \emph{et al.}  reproduce the failure of network reciprocity~\cite{Ezaki2017Reinforcement} and discuss the condition for cooperation emergence~\cite{Masuda2011Numerical}. Jia \emph{et al.} use the Monte Carlo method to study pattern formation and phase transitions towards cooperation in social dilemmas that are driven by RL, they show that global players play a decisive role in ensuring cooperation~\cite{Jia2021Local}. Song \emph{et al.} combine the iterated prisoner’s dilemma game with the Bush-Mosteller reinforcement learning model and show that there exists a moderate switching dynamics of the interaction intensity that is optimal for the evolution of cooperation~\cite{Song2022Reinforcement}. In Ref.~\cite{Wang2022Levy}, when reinforcement learners playing PD game are subject to L\'{e}vy noise, a particularly positive role of L\'{e}vy noise is unveiled within this social dilemma. Ding \emph{et al.} reveal that a strong memory and a long-sighted expectation yield the emergence of coordinated optimal policies, where both players act like ``win stay, lose shift" to maintain a high level of cooperation~\cite{Ding2023Emergence}.

Till now, these studies are primarily based on the PD where only two players are engaged in each single game. 
Only recently, few works have started to discuss the evolution of multiplayer game within the RL paradigm by adopting PGG. In Ref.~\cite{Dan2022Empty}, the researchers investigate how cooperative behavior evolves in a PGG on a network with empty nodes based on the Bush-Mosteller model; the results show that only the existence of appropriate empty nodes in the network can stimulate individual cooperative behavior. Wang \emph{et al.} propose a framework that combines the PGG and adaptive reward mechanisms to better capture decision-making behaviors in multi-player interactions, where they are mainly concentrating on the synergistic effects of them in RL~\cite{Wang2023Synergistic}. Ref.~\cite{Zhang2024Exploring} introduces the Q-learning algorithm into the voluntary PGG to explore the evolution of cooperation, and uncovers when the synergy factor is large and the adjust loner payoff’s multiply factor is smaller, the number of cooperators becomes gradually consistent, and the proportion of defectors evolves non-linearly for different synergy factors.

However, these works based on PGG consider external factors, such as empty nodes or rewards, it's unclear what's evolution of cooperation within the original PGG game. More importantly, these works only consider the individual's own action information or are termed as a self-regarding setup, where the players neglect the information of surroundings. This setup simplifies the model but obviously contradicts our daily experiences, where our decision-making is based upon not only our own actions but also the environmental information (e.g. neighbors' actions) we perceived. It is generally believed that once more information about their surroundings is incorporated, we could make better decisions to maximize the payoffs\textcolor{blue}{~\cite{Sutton2018reinforcement}}.

In this work, we investigate the evolution of cooperation of the original PGG within the paradigm of RL where the environmental information is incorporated.
Specifically, we empower players with the Q-learning algorithm to revise their strategies, where each player is guided by a Q-table.
The environmental information is incorporated into the state of the Q-table in a way that keeps the Q-table simple enough. Our simulation shows that the incorporation of environmental information is able to promote the cooperation prevalence, and cooperation is more likely to emerge than the case of IL.
Besides, for the PGG with voluntary participation, the cyclic dominance in the previous studies~\cite{Phase2002Szabo} within IL disappears; instead, a non-monotonic cooperation dependence of the game parameter is observed, which is counter-intuitive.


\begin{figure*}[htbp!]
\centering
\includegraphics[width=0.8\textwidth]{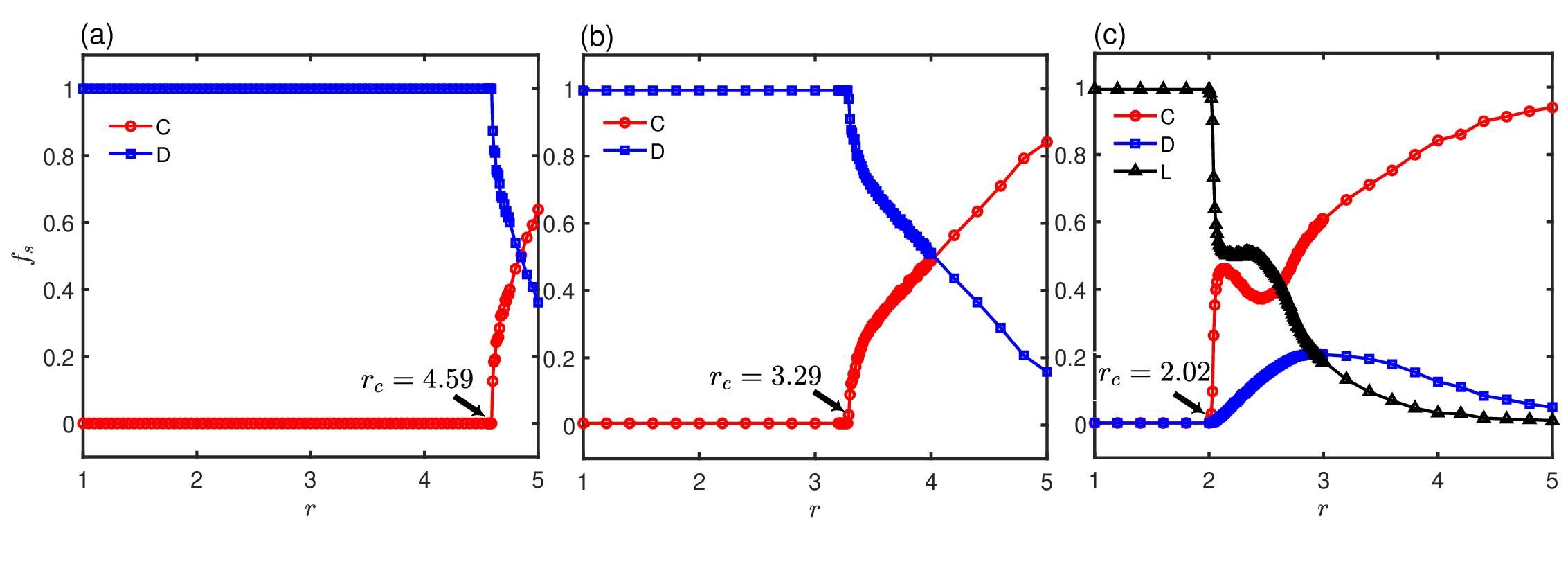}
\caption{(Color online) {\bf Cooperation phase transition versus the gain factor $r$ for three different models.}
(a) Players update their strategies using the Fermi rule in PGG with $K$=0.1, and the threshold for cooperation emergence is around $r_c$=4.59.	
(b) Players update their strategies using the Q-learning algorithm in PGG, and the threshold for cooperation emergence is around $r_c$=3.29.
(c) Players update their strategies using the Q-learning algorithm in VPGG, and the threshold for cooperation emergence is around $r_c$=2.02.
Other parameters: $\epsilon$=0.01, $\alpha$=0.1, $\gamma$=0.9, $\sigma$=1.0, and the population size $N=100\times100$.
} 
\label{fig:PT} 
\end{figure*}
\section{2. Model} 
\label{sec:model}

Let's consider a square lattice with periodic boundary condition, where each node represents an individual engaging in the public goods game. Specifically, we investigate two models with the Q-learning algorithm: the public goods game (PGG) and the voluntary public goods game (VPGG). In the first model, the focal player initiates a public goods game with four of its nearest neighbors, where each player can either choose cooperation (C) or defection (D). The choice of C means that an individual invests a certain amount (set to 1) in the public pool, while D represents a free-riding behavior without any contribution. Different from PGG, participation in VPGG is voluntary, players can choose to be ``loner" (L) as the third strategy, if they do not want to play the game at all, and receive a small fixed income denoted as $\sigma$ (fixed at 1 if not stated otherwise). It is noteworthy that Ref.~\cite{Zhang2024Exploring} mainly focus on the impact of the loner strategy with self-regarding setup on the evolution of cooperation, without considering the environmental information of individuals and using adjustable payoffs for loners. Here, we maintain consistency with previous studies on IL~\cite{Phase2002Szabo} by fixing the loner's payoff at 1, while incorporating individuals' environmental information into the Q-table setup to study the evolution of cooperation within the RL framework. This allows for a better comparison with the results of IL.

Suppose now a player $i$ initiates a PGG, five participants ($n_i=5$) are involved, while the number of participants in VPGG needs to subtract the number of loners $n_L$ in their neighborhood. After the investment, the amount of money in the public pool is multiplied by a gain/synergy factor $r$, and then evenly divided among all participating players. The expected payoffs for different types of players $i$ are
\begin{equation}
\pi_i= \begin{cases}\frac{r n_C}{n_C+n_D}-1, & \text { if } s_i=C \\
                               \frac{r n_C}{n_C+n_D}, & \text { if } s_i=D \\ 
                               \sigma, & \text { if } s_i=L
                               \end{cases}
\label{eq:payoff}
\end{equation}
where $s_i$ represents the strategy of the player $i$, $n_{C,D}$ represents the number of cooperators and defectors, respectively. By convention, in the case of $r\in(1,n_i)$ or $r\in(1,n_i - n_L)$, social dilemmas arise where the choice D brings a higher payoff than C.

In our models, individuals update their strategies according to the Q-learning algorithm~\cite{Watkins1989learning,Watkins1992Q}, where each has a Q-table to guide their decision-making. 
The Q-table is a two-dimensional table expanded by states and actions.
The action set consists of all available strategies, which is $\mathbb{A}= \{$C$, $D$\}$ in PGG, and $\mathbb{A}= \{$C$, $D$, $L$\}$ in VPGG.
The state set serves to characterize the environment perceived by individuals. In the previous self-regarding setup, the state only records the action adopted by the player itself in the last round.
Here, the environment information is captured in a simple manner by comparing the number of cooperators $n_C$ and defectors $n_D$ among the player's four nearest neighbors. Therefore, there are three possible states $s_{1,2,3}$ defined by the condition of $n_C\textgreater n_D$, $n_C=n_D$, $n_C\textless n_D$, respectively. We use the same state set in both models $\mathbb{S}= \{s_1, s_2, s_3\}$. The two Q-tables are shown at the bottom of Fig.~\ref{fig:model}. The idea of Q-table is score to the value of each action $a$ within each state $s$. A larger value of $Q_{s,a}$ within a given row is supposed to be more valuable within that state, which means the action $a$ is more likely to be chosen when within the state $s$.


\begin{algorithm}
\label{algorithm:1}
\SetAlgoNlRelativeSize{0} 
\SetInd{0.5em}{0.5em} 
\SetAlgoNlRelativeSize{-1} 
\SetNlSty{text}{}{\hspace{0.5em}} 
\caption{Pseudocode of PGG/VPGG with Q-learning}
\KwIn{$\alpha,\gamma,\epsilon$}
\textbf{Initialization}\;
\For{each agent}{
    Pick an action randomly from $\mathbb{A}$\;
    Create a Q-table with each item in the matrix near zero\;
}
\For{each agent}{
    Generate state: $s$\;
}
\Repeat{the termination condition is met}{
    \For{each agent}{
        \eIf{rand() $< \epsilon$}{
            Pick an action randomly from $\mathbb{A}$;
        }{
            Takes action according to the Q-table\;
            $a = \arg \max_{a_i} Q_{s, a_i}$\;
        }
    }
    \For{each agent}{
        Compute rewards according to Eq.(\ref{eq:payoff})\;
        Update state and Q-table according to Eq.(\ref{eq:Q-learning})\;
    }
}
\end{algorithm}

Without loss of generality, each player is randomly assigned with an action from the action set $\mathbb{A}$, and the elements $Q_{s,a}$ in the Q-tables randomly assigned a value between [0, 1) independently. 
In our Monte Carlo simulations, the evolution follows a synchronous updating procedure.
In each round $t$, with a probability $\epsilon$, they opt for a random action $a\in\mathbb{A}$; with probability $1-\epsilon$, they strictly follow the guidance of their Q-tables by choosing the action with the largest Q-value given their current states. After all players make their moves, they receive the reward $\pi_i$ following Eq.~(\ref{eq:payoff}), and are able to compute the new state $s'$. As the next step, each player $i$ tries to draw lessons by revising the corresponding $Q^i_{s, a}$ they just adopted in this round, according to the Bellman equation~\cite{Sutton2018reinforcement}:
\begin{equation}
\begin{aligned}
Q^i_{s, a}(t+1) 
&=(1-\alpha) Q^i_{s, a}(t)+\alpha\left(\pi_i(t)+\gamma \max _{a^{\prime}} Q^i_{s^{\prime}, a^{\prime}}(t)\right),
\end{aligned}
\label{eq:Q-learning}
\end{equation}
where $s$ and $a$ represent the current state and action of the focal individual $i$, and $s^{\prime}$ is the new state at $t+1$. $\alpha\in(0,1]$ is the learning rate determining how fast of the Q-value is revised, $\gamma\in[0,1]$ is the discount factor, defining the significance of future rewards. Notice that, $\pi_i(t)$ is the payoff for player $i$ when she/he acts as the focal player initiating a PGG game, not all payoffs received. Because the item $Q^i_{s, a}$ reflects the value of action the player $i$ actively made\textcolor{blue}{~\cite{Zhang2020Oscillatory}}.
When all players update their Q-tables, a single MC round is completed. 
The evolution consists of many MC steps and is terminated until the system reaches equilibrium or the desired duration is reached, the pseudocode is provided in Algorithm~\ref{algorithm:1}.

Besides the two models using Q-learning proposed above, we also adopt the Fermi updating rule~\cite{Szabo1998Evolutionary} in the traditional framework of IL as the baseline model for comparison.  Here, players make decisions by comparing utility with their neighbors and imitate strategies of those who are better off in the game. Since this setup has been extensively studied in previous game-theoretic works, we refer to Ref.~\cite{Phase2002Szabo,Lucas2022Cooperation} for the simulation details. The three involved model setups are intuitively illustrated in Fig.~\ref{fig:model}.

\begin{figure*}[hpb!]
\centering
\includegraphics[width=0.8\textwidth]{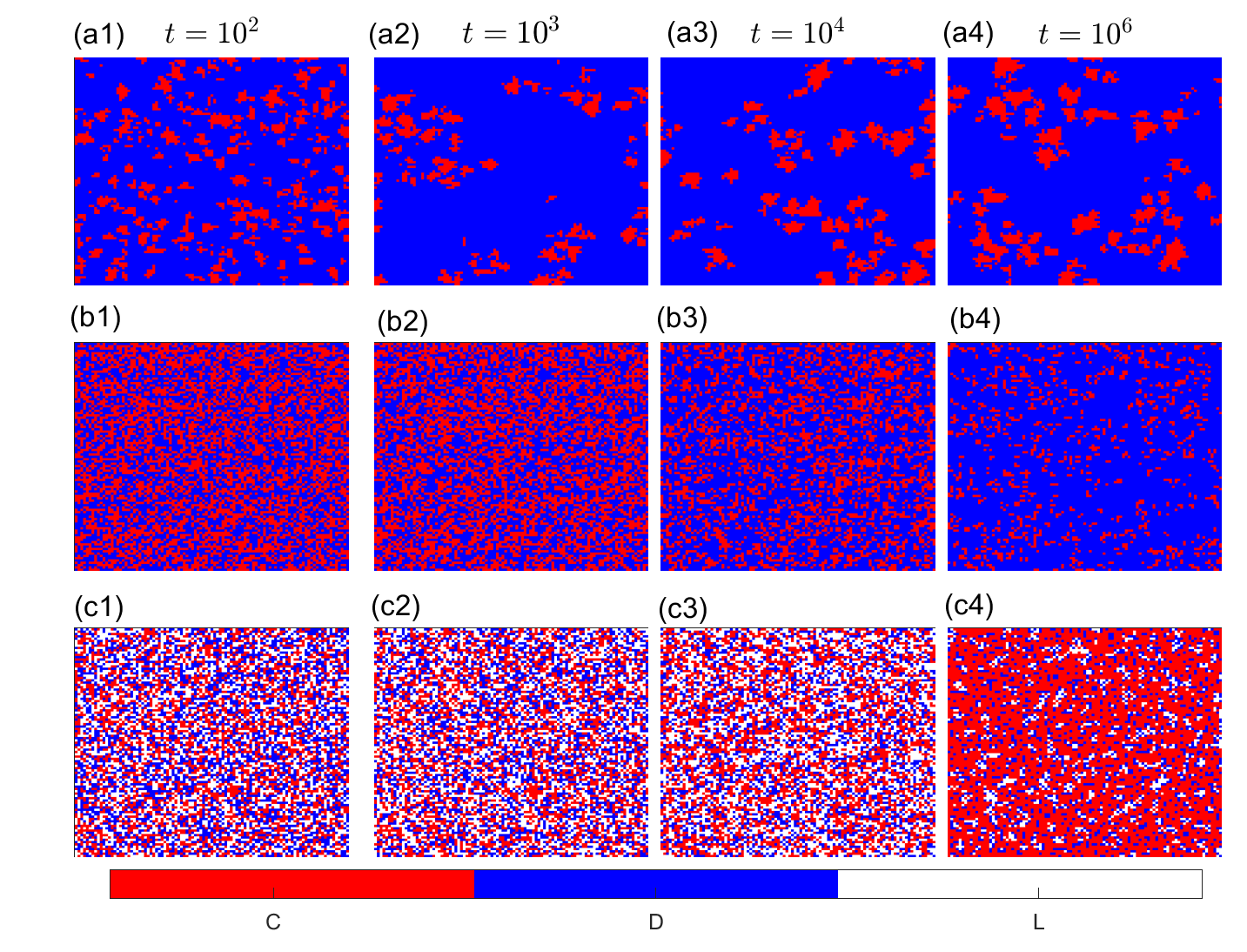}
\caption{(Color online) {\bf Typical cooperation snapshots for the three setups.} 
In (a1-a4), PGG with Fermi rule, and cooperation clusters are formed to resist the exploitation by defectors. The gain factor $r=4.60$.
In (b1-b4), PGG with the Q-learning algorithm, no apparent clusters of cooperators are seen instead, and the evolution time takes longer. 
In (c1-c4), VPGG with the Q-learning algorithm, there are also no obvious cooperation clusters, but the resulting cooperation prevalence $f_c$ in panel (c4) is much higher than the panel (b4).
 $\epsilon =0.01$, $\alpha$=0.1, $\gamma$=0.9, $\sigma$=1.0, $r=3.32$ are used in both (b1-b4) and (c1-c4).
 Other parameters: the population size $N=100\times100$.
} 
\label{fig:pattern} 
\end{figure*}

\begin{figure*}[htpb!]
\centering
\includegraphics[width=0.8\textwidth]{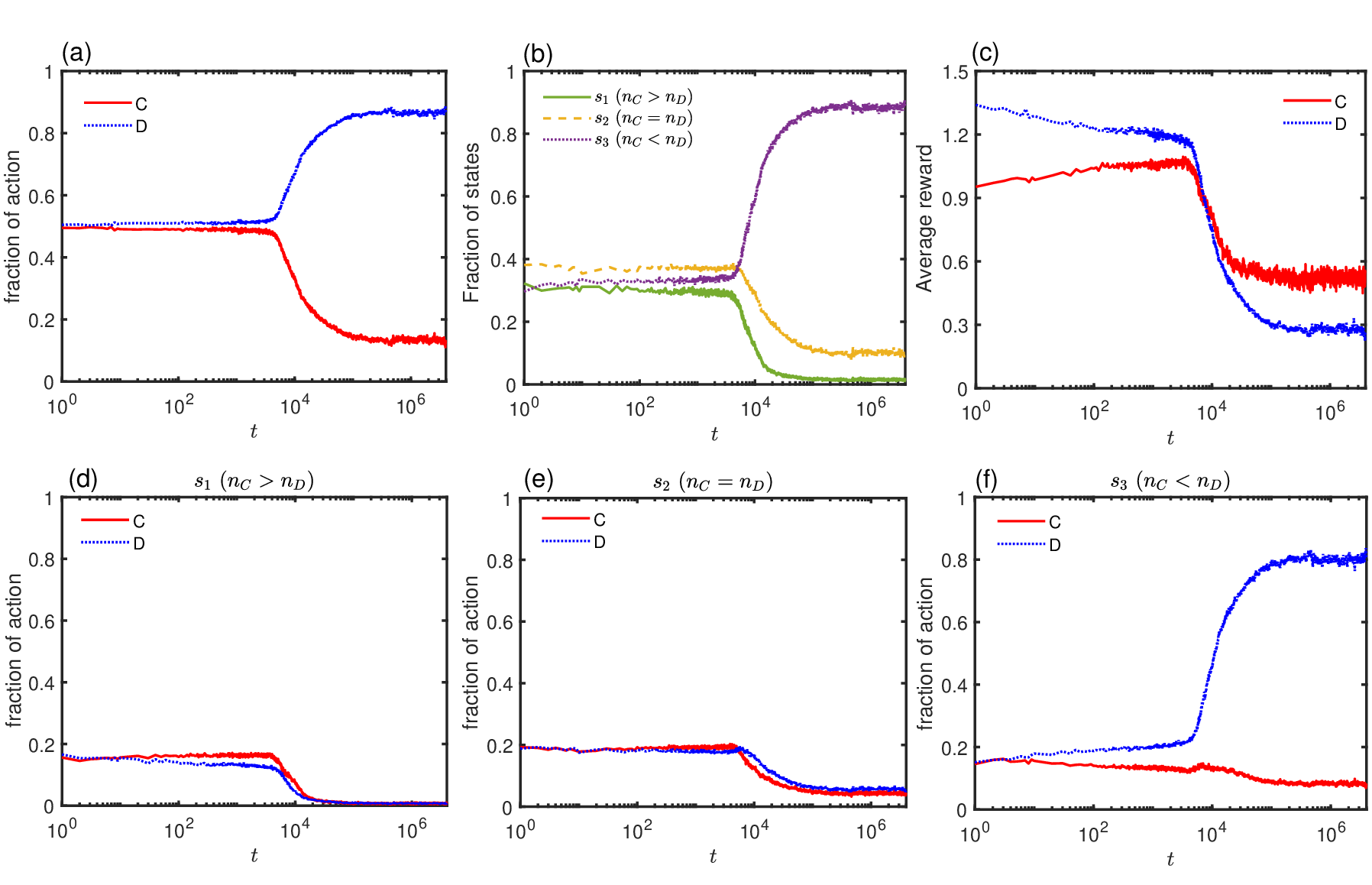}
\caption{(Color online) {\bf Typical time series in PGG with Q-learning}.
(a) Time evolution of two strategies, which differs from the typical ``first down and later up" trend often observed as a sign of network reciprocity in the framework of imitation learning.
(b) The temporal evolution of the fractions of the three states $s_{1,2,3}$. After the transient period, $s_3$ dominates.
(c) The evolution of average reward for two different strategies, where there is a crossover observed.
(d-f) Time evolution of action preference in three different states. The emergence of cooperation primarily occurs in state $s_1$ ($n_C<n_D$).
Other parameters: $\epsilon =0.01$, $\alpha$=0.1, $\gamma$=0.9, $\sigma$=1.0, $r$=3.32, and the population size $N=100\times100$.
} 
\label{fig:mechanismFPGG} 
\end{figure*}

\section{3. Results}\label{sec:results}

We first report the fraction of different strategies as a function of the gain factor $r$ in Fig.~\ref{fig:PT} for the three models. 
Fig.~\ref{fig:PT}(a) shows the results with the Fermi updating rule, where we observe that cooperation emerges as the gain factor is larger than a critical threshold $r>r_c\approx 4.59$, in line with previous studies~\cite{Lucas2022Cooperation}. When the gain factor $r$ is high enough, a cooperation phase transition is seen as shown here. Notice that, the model with a strictly synchronous updating exhibits similar phase transition, with the threshold being $r_c\approx 4.80$ (data now shown), showing a bit more difficult for the cooperation to emerge.
Fig.~\ref{fig:PT}(b) shows the results for PGG with the Q-learning algorithm. Compared to Fig.~\ref{fig:PT}(a), the cooperation phase transition is qualitatively the same, but with a reduced threshold $r_c\approx 3.29$. This means that cooperation is more likely to emerge within the reinforcement learning paradigm than in the case of imitation learning.

When voluntary participation is introduced, Fig.~\ref{fig:PT}(c) shows that the threshold is further reduced to $r_c\approx 2.02$ where cooperation is most likely to emerge among the three models we studied here. Detailed examination indicates that the introduction of ``loner" is an efficient manner to considerably inhibit the growth of defectors. For a lower gain factor $r<r_c$,  the fraction of loners dominate. Defectors appear only when cooperators survive as $r>r_c$, but as the gain factor further increases cooperators dominates. The fraction of defectors remains $f_D\lesssim0.2$ for the whole range. This overall picture aligns with findings from previous work~\cite{Phase2002Szabo} by IL.  However, there is a non-monotonic dependence region for $r\in(2.13, 2.50)$ where the increase in $r$ leads to the reduction in $f_C$ accompanied by the increase in $f_D$. Notice that, the non-vanishing $f_L$ at the end of $r\rightarrow 5$ effectively inhibits the invasion of defectors, guaranteeing a higher level of cooperation compared to the case without loner [see Fig.~\ref{fig:PT}(b)].
A systematic investigation of the impact of the two learning parameters is given in Appendix A. It shows that when individuals both focus on historical experience and have a long-term vision, cooperation is likely to emerge.  Otherwise, either a forgetful property (a large $\alpha$) and/or a short-term vision (a small $\gamma$) lead to the failure of its emergence.

To develop some intuition for the evolution process for the three scenarios, we present some typical spatio-temporal snapshots of the system.
Starting from random initial conditions, Figs.~\ref{fig:pattern}(a1-a4) show that some cooperation clusters are formed. This is a typical illustration of network reciprocity~\cite{Nowak1992Evolutionary}, where the structured population enables cooperators to persist by forming clusters, thereby avoiding the exploitation by defectors. For the given parameter $r=4.60$, being slightly larger than the threshold, the resulting prevalence of cooperation is stabilized around 0.13, while the cooperation clusters are not frozen but fluctuate all the time.

The PGG with the Q-learning algorithm is illustrated in Fig.~\ref{fig:pattern}(b1-b4), where $r$ is also chosen slightly larger than the threshold. As can be seen, it's evident that the patterns are quantitatively distinct from Figs.~\ref{fig:pattern}(a1-a4), where there are no apparent cooperator clusters observed. Over time, the number of cooperators gradually declines, and upon reaching a steady state, these cooperators are evenly scattered across the whole domain. Additionally, the time-scale reaching a steady state seems longer in this case than in Figs.~\ref{fig:pattern}(a1-a4).

Finally, we showcase the pattern evolution for the VPGG with Q-learning, with identical parameter $r = 3.32$ as adopted in Fig.~\ref{fig:pattern}(b1-b4) for comparison. Similar to the PGG, no obvious cooperation clusters are seen in Fig.~\ref{fig:pattern}(c1-c4). As time goes by, however, both the numbers of loners and defectors decrease, and cooperators dominate in the end, with a notably higher cooperation prevalence compared to Fig.~\ref{fig:pattern}(b1-b4). Furthermore, the patterns depicted do not display the waves with cyclic dominance as observed in IL~\cite{Phase2002Szabo}. This suggests a different mechanism working here.

\begin{figure}[bpth]
\centering
\includegraphics[width=0.8\linewidth]{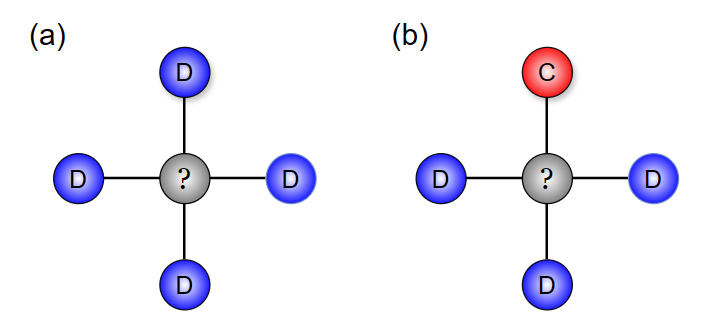}
\caption{(Color online) {\bf Two scenarios within state $s_3$ where $n_C<n_D$.}
The gray circle represents the focal individual, who is either to cooperate or to defect.
Red and blue circles represent the cooperators and defectors in neighbors, respectively. 
Cooperation is almost impossible in scenario (a), while it may occur in scenario (b).
}
\label{fig:TwoScenarios}
\end{figure}

\section{4. Mechanism analysis}\label{sec:mechansim}

To understand how cooperation evolves when people play PGG with Q-learning algorithm. In this section, we move to the mechanism analysis of the latter two cases playing with Q-learning.

\subsection{4.1 Mechanism analysis of PGG with Q-learning}\label{sec:two}

We commence our analysis by delving into PGG with Q-learning, where the cooperation phase transition and pattern evolution have been shown above. 
To further observe the temporal evolution, we present time series in Fig.~\ref{fig:mechanismFPGG}(a), with identical parameters adopted in Fig.~\ref{fig:pattern}(b1- b4). Two strategy fractions remain at the initial level around 0.5 and begin to decline later on and eventually stabilize. This observation is fundamentally different from the ``first down and later up" trend often regarded as a sign of network reciprocity in IL~\cite{Perc2008Restricted,Szolnoki2009Promoting}.

To understand this trend, we need to take into account the surrounding information and the resulting payoffs.
As depicted in Fig.~\ref{fig:mechanismFPGG}(b), with random initial conditions, the proportions of the three states $s_{1,2,3}$ are approximately identical. Fig.~\ref{fig:mechanismFPGG}(c) shows that initially free-riders (D) consistently yield higher payoffs than cooperators (C) since approximately half of the individuals choose to cooperate. As a result, players tend to defect by learning, leading to an increasing proportion of the state $s_3$ ($n_C < n_D$), and accordingly the other two state fractions of $s_{1,2}$ decrease. 
Specifically, the state $s_3$ includes two different scenarios (see Fig.~\ref{fig:TwoScenarios}):
(a) The focal individual is surrounded by four defectors, in which case choosing C yields a negative payoff, which is disadvantageous.
(b) There is one cooperator present in the neighborhood around the focal individual, choosing C will result in a positive payoff, leading to an increase in the Q-value in this state. While opting for strategy D yields a higher payoff but this also regresses the environment back to scenario (a).

\begin{figure}[htbp!]
\centering
\includegraphics[width=0.8\linewidth]{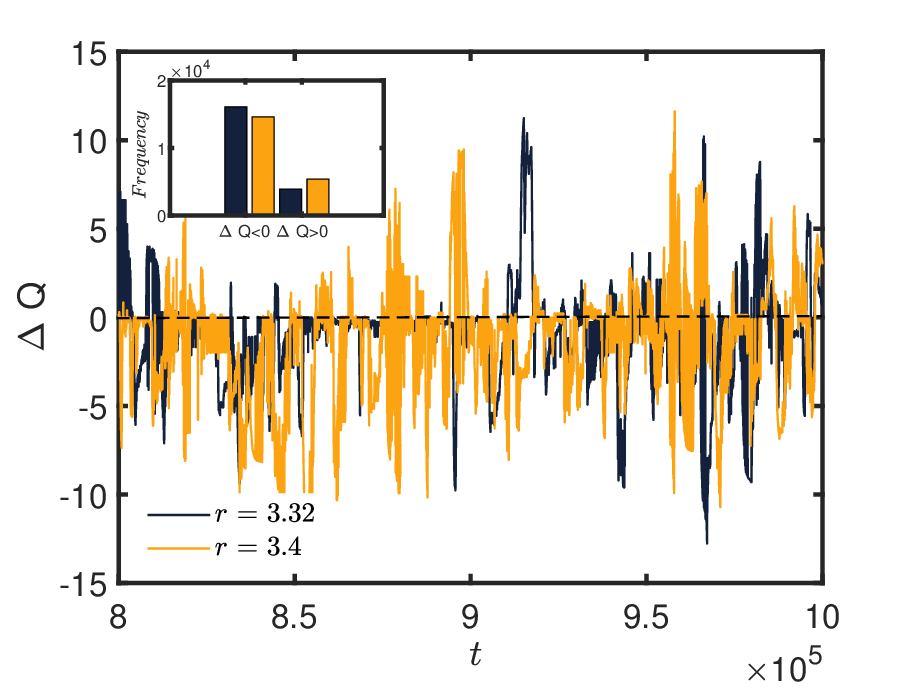}
\caption{(Color online) {\bf Time evolution of $\Delta$Q after transient within state $s_3$.} 
Typical time evolution of $\Delta Q=Q_{s_3,C}-Q_{s_3,D}$ in state $s_3$ ($n_C\textless n_D$). The inset shows the frequencies of occurrences when $\Delta Q >0$ and $\Delta Q <0$ at the same time intervals. Two parameters $r=3.32$ (black) and $r=3.4$ (orange) are chosen here for comparison.
 Other parameters: $\epsilon=0.01$, $\alpha=0.1$, $\gamma=0.9$, and the population size $N=100\times100$.
} 
\label{fig:DeltaQ} 
\end{figure}

In the process of decision-making using the Q-learning algorithm, individuals focus on both historical experience and long-term rewards, therefore they try to avoid falling into scenario (a). Consequently, by factoring in the low payoff in scenario (a) and the potential for higher payoffs in the new state $s_2$ ($n_C=n_D$), the preference of choosing C in state $s_3$ arises, leading to the emergence of cooperation. In fact, the choice of C indeed brings higher payoffs on average as shown by the crossover in Fig.~\ref{fig:mechanismFPGG}(c). In Fig.~\ref{fig:mechanismFPGG}(d-e), we show the evolution of action preferences in three different states. It can be observed that the emergence of cooperation primarily occurs mostly in the state $s_3$.

\begin{figure*}[htbp!]
\centering
\includegraphics[width=0.8\textwidth]{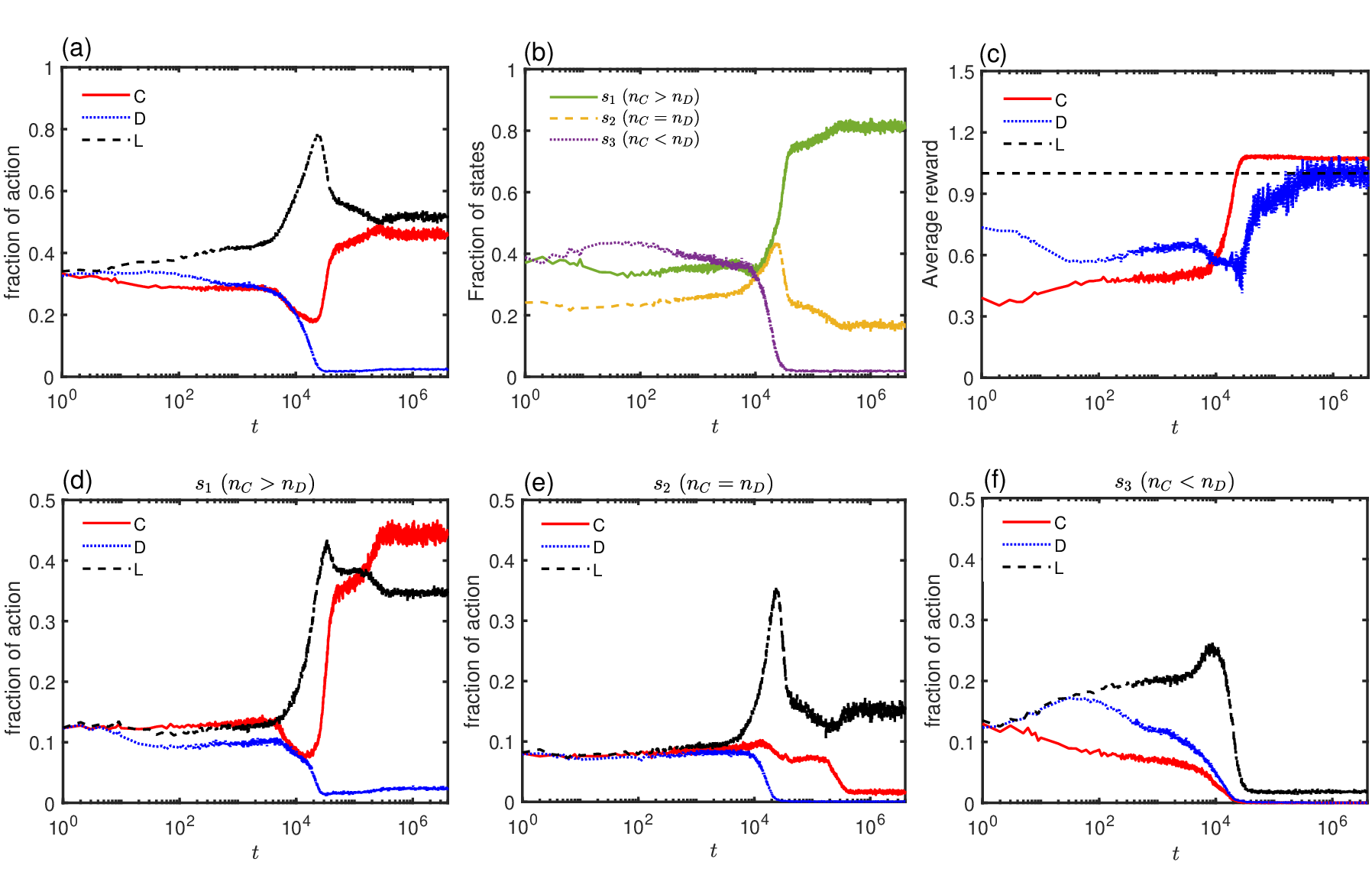}
\caption{(Color online) {\bf Typical time series in VPGG with Q-learning.} 
(a) Time evolution of three strategies. As the defectors begin to decline, the fraction of cooperators shifts from a decreasing to an increasing trend, while the corresponding loners shift from an increasing to a decreasing trend, indicating the emergence of cooperators within clusters of loners.
(b) Time evolution of the fractions of the three states. After a sufficiently large number of Monte Carlo steps, it can be observed that the system is most likely in the state of $s_1$, while there exists a certain proportion of states $s_2$ and $s_1$ in dynamic equilibrium.
(c) The evolution of average reward for three different strategies. In the early stages, loners consistently generate higher average payoffs compared to the other strategies.
(d-f) Time evolution of action preference in three states $s_{1,2,3}$. The emergence of cooperation primarily occurs in state $s_1$.
 Other parameters: $\epsilon=0.01$, $\alpha=0.1$, $\gamma=0.9$, $r=2.11$ and the population size $N=100\times100$.
} 
\label{fig:VFPGG} 
\end{figure*}

To further confirm, we monitor the preference evolution captured by the Q-values of a typical individual within state $s_3$, as shown in Fig.~\ref{fig:DeltaQ}. It's convenient to define $\Delta Q=Q_C- Q_D$ as the cooperation preference, a positive value corresponds to the case that cooperation is preferred by learning and vice versa. With this in mind, we observe that the value of $\Delta Q$ fluctuates up and down around 0. Most of the time, $\Delta Q<0$ with intermittent occurrences of $\Delta Q>0$. This means that D is the preferred mostly but C is also selected from time to time. The two actions compete and reach a balance as seen in Fig.~\ref{fig:mechanismFPGG}(f).
 As the gain factor $r$ increases, the payoff advantage of defectors over cooperators diminishes, indicating an increased likelihood of cooperation emergence. This is validated in the inset of Fig.~\ref{fig:DeltaQ}, which shows an increased frequency of $\Delta Q>0$ when $r$ increases to 3.4.

The above analysis indicates that self-reflective decision-making within Q-learning does not rely on clusters of cooperators to resist defectors. Instead, it possesses the potential to escape from the most adverse environments of full defection, thereby leading to the emergence of cooperation.

\subsection{4.2 Mechanism analysis of VPGG with Q-learning}\label{sec:three}

We next turn to the mechanism in the VPGG with the Q-learning, especially the impact of voluntary participation, shown in Fig.~\ref{fig:VFPGG}. 
Starting from random initial conditions, the fraction of loners (L) starts to increase, and both defectors and cooperators decrease. But once the fraction of D stabilizes near zero, cooperators start to increase and loners decrease, and eventually, the two fractions become stabilized with approximately equal size for the given parameters, shown in Fig.~\ref{fig:VFPGG}(a). This evolution is also captured by the fraction of the three states $s_{1,2,3}$, where the fraction of state $s_1$ dominates and the fraction of $s_3$ diminishes.

Fig.~\ref{fig:VFPGG}(c) shows that the payoff of the loner is much larger than the other two fractions at the initial stage, which explains the number increases in loners. Figs.~\ref{fig:VFPGG}(d-f), respectively, show the evolution of action fraction within the three states $s_{1,2,3}$,  where the action preference differs significantly. To specifically understand the non-trivial evolution trend in Fig.~\ref{fig:VFPGG}(a), we need to examine the action prevalence in the three states $s_{1,2,3}$ by incorporating the state fractions shown in Fig.~\ref{fig:VFPGG}(b). At the first stage ($t\lesssim 10^4$), the state fraction of $s_3$ is slightly higher than the other two, where the action fraction of loner is highest ($10^2\lesssim t\lesssim 10^4$), the three action fractions in $s_{1,2}$ are approximately identical; these observations then explain the increasing trend in loners and decreasing trend of both C and D.

At around $t=10^4$, the action fraction of L becomes largest in all three $s_{1,2,3}$, and loner becomes the most preferred choice in all cases by learning at this moment. This change then leads to the peak of fraction in L in Fig.~\ref{fig:VFPGG}(a). As $f_D\rightarrow 0$, this change provides an ideal surrounding for players to choose cooperation, which brings higher rewards than the case of being loners (see Fig.~\ref{fig:VFPGG}(c)). Therefore, the state $s_1$ dominates afterward, and the action of C becomes a more preferred action than L, which is supported by the payoff comparison provided in Table I in Appendix B. That's the reason for the crossover for loners and cooperators as seen in later evolution in Fig.~\ref{fig:VFPGG}(d).
Though, within state $s_2$, being a loner is a better choice than being a cooperator (see Table II in Appendix B). Within state $s_3$, being a loner is still the best choice, and being a cooperator is worst, though the fraction of $s_3$ is too small.  

Taken together Fig.~\ref{fig:VFPGG}(a-c), players opt to be loners within $s_{2,3}$, and cooperator is more preferred within $s_1$. This is the reason why only cooperators and loners survive and coexist.
The overall picture is that individuals in this setup choose strategy L from time to time to inhibit the growth of defectors, laying the groundwork for cooperators to thrive in a sea of loners.

\begin{figure}[htbp!]
\centering
\includegraphics[width=0.8\linewidth]{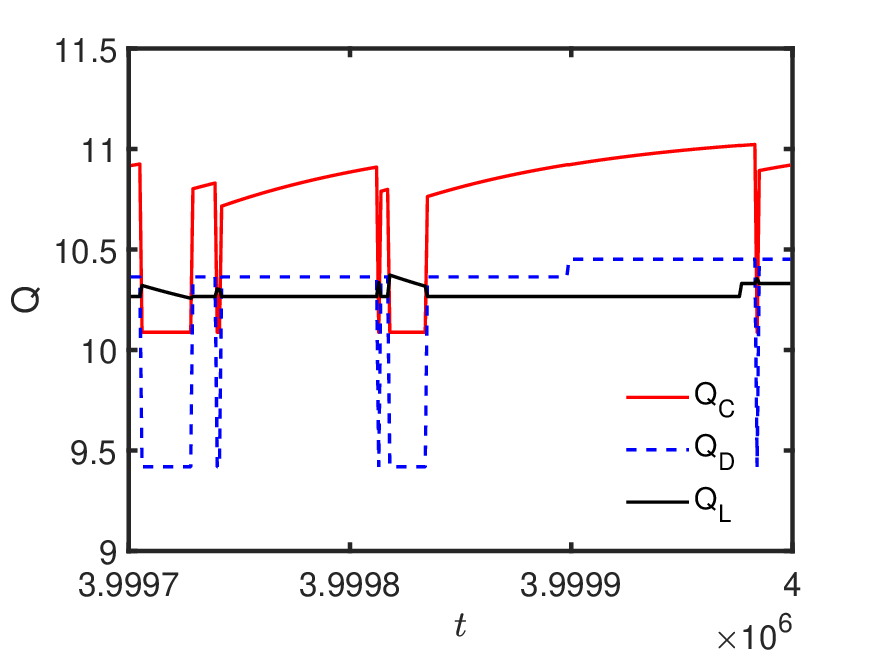}
\caption{(Color online) {\bf Time evolution of Q value in a stable state.} 
In VPGG, time series of Q-value for three different strategies in a stable state.
 Other parameters: $\epsilon=0.01$, $\alpha=0.1$, $\gamma=0.9$, and the population size $N=100\times100$.
} 
\label{fig:DeltaQStable} 
\end{figure}

By this point, one might question are the coexistence stability of L and C, and whether the strategy L can entirely be transformed into C or vice versa. The answer is no. An all-C system is unstable and susceptible to the invasion of D. Similarly, an all-L system is also unstable and can be invaded by C, and an all-D system is susceptible to the invasion of L. This cyclic dominance determines that none could dominate, they have to coexist in the form of a subtle balance, even though the fraction of D is very small. 
An example is shown in Fig.~\ref{fig:DeltaQStable}, where we present the time series of Q values for all three strategies in a stable state, randomly selecting an individual. We see that $Q_{C}$ fluctuates up and down around $Q_{L}$, while $Q_{D}$ shows a dependence on $Q_{C}$, their orders are reshuffled consistently.
This observation demonstrates that the three strategies are in dynamic equilibrium.

\subsection{4.3 Analysis of non-monotonic variation}\label{subsec:normal_scenarios}

\begin{figure}[htbp!]
\centering
\includegraphics[width=0.8\linewidth]{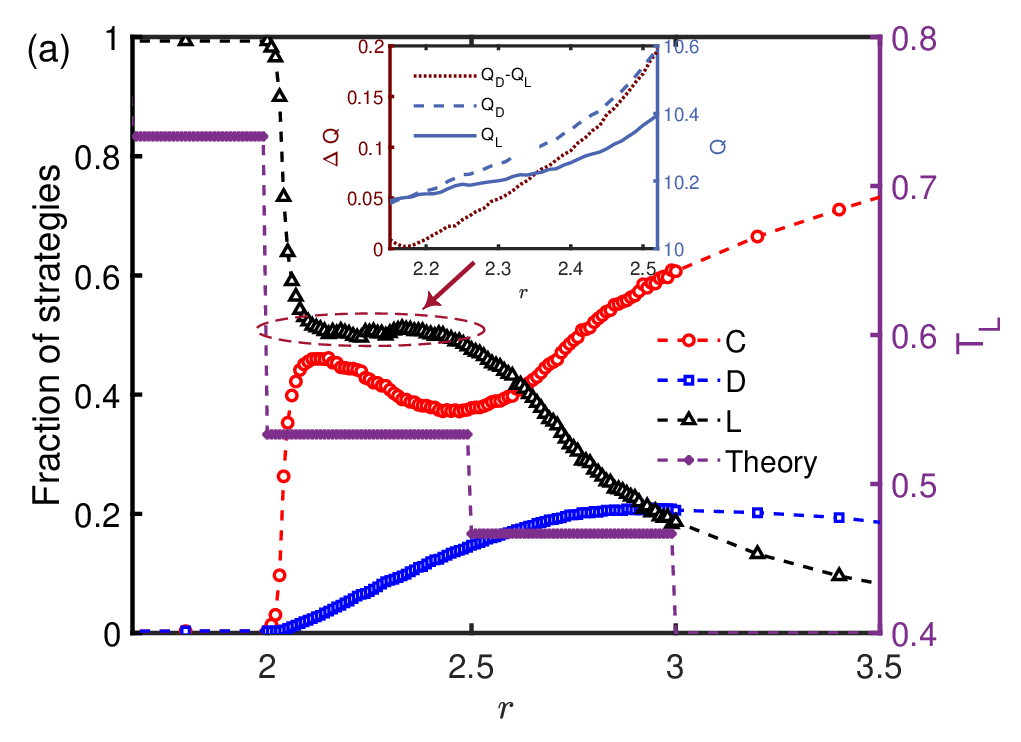}
\includegraphics[width=0.8\linewidth]{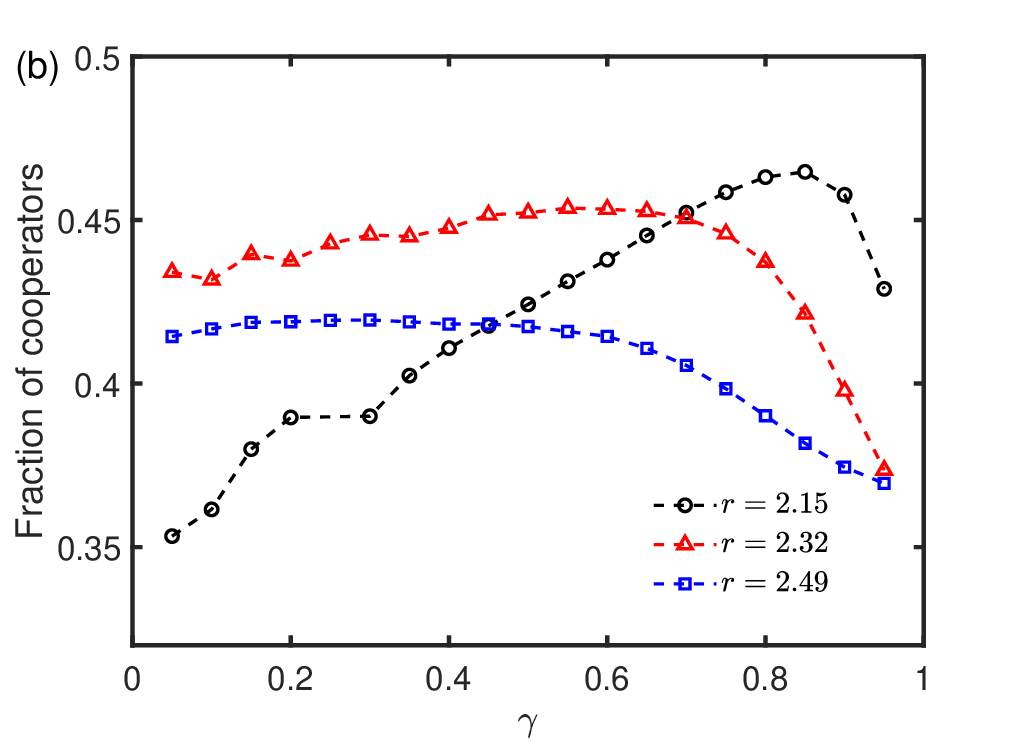}
\caption{(Color online) {\bf The analysis of non-monotonic interval in VPGG.} 
(a) The left side of the y-axis shows the fraction of strategies as a function of the gain factor $r$. 
The right side shows the tendency of choosing strategy L based on theoretical analysis, characterized by the proportion of players where both conditions of $\pi_L>\pi_C$ and $\pi_L>\pi_D$ are satisfied. 
The inset shows the corresponding values of $\Delta Q$ and $Q_{L,D}$. 
(b) The fraction of cooperators as the discount factor $\gamma$ varies for three different $r$ values.
 Other parameters: $\epsilon =0.01$, $\alpha=0.1$, $\gamma = 0.9$, and the population size $N=100\times100$.
} 
\label{fig:NonMonotonic} 
\end{figure}
Till now, the non-monotonic dependence of cooperation level for $r\in(2.13, 2.50)$ in Fig.~\ref{fig:PT}(c) remains unexplained. In this interval, the fraction of the strategy L remains nearly constant around 0.5, and the cooperation prevalence decreases as the gain factor $r$ is increased.

To proceed, we estimate the tendency to choose strategy L in theory by computing the proportion of players where the condition $\pi_C \textless \pi_L$ and $\pi_D \textless \pi_L$ are both satisfied within all possible states. Here, $\pi_C$, $\pi_D$, and $\pi_L$ represent the payoffs of choosing C, D, and L, respectively. This reveals the tendency of individuals to choose the strategy L as a function of $r$, as depicted in Fig.~\ref{fig:NonMonotonic}(a).
As shown in Fig.~\ref{fig:NonMonotonic}(a), the tendency to choose strategy L exhibits a stage-wise plateau in the interval $r\in(2.15,2.49)$. Within this range, the fraction of loners remains fixed, while the fraction of cooperators shows a decreasing trend. As $r$ further increases to $r\geq2.5$, the tendency of loners continues to decrease, approaching a new plateau, from the results of numerical simulations, we notice a reduction in the fraction of loners and a simultaneous increase in the fraction of cooperators. Notice that, these stage-wise plateaus correspond to the present rewards that are immediately received at present, where the impact of expected payoffs in the future is not included.

The reason why the plateau well matches the non-monotonic interval is exactly because the impact of expected payoffs in the future is weakened. 
As depicted in Fig.~\ref{fig:NonMonotonic}(b), three typical cases within the non-monotonic region $r\in(2.13, 2.50)$ are shown, where we observe that the dependence of the cooperation prevalence on the discount factor $\gamma$ is non-monotonic, there is an optimal value of $\gamma$ in each case yielding the highest cooperation prevalence.  
This is in sharp contrast with the results obtained from previous observations~\cite{Ding2023Emergence, Zheng2024decoding}, where a higher expectation (i.e. a larger $\gamma$) always leads to a better outcome of evolution; two such examples outside of the non-monotonic region ($r=2.05$ and 2.66) are shown in Fig.~\ref{fig:Deviate_nonmonotonic} in Appendix C~\ref{sec:appendixC}. This difference implies that, within this non-monotonic interval, a high emphasis on future rewards is less favorable for the emergence of cooperation, suggesting potential challenges in maintaining long-term cooperation stability, the underlying cause of this result is the uncertainty in the environment due to changes in the strategies of other individuals.

Specifically, as discussed in the above section, L and C are more preferred than D within the three states $s_{1,2,3}$ in typical scenarios. 
But this is not always the case, the inset in Fig.~\ref{fig:NonMonotonic}(a) shows that the Q-value for L is no longer larger than the value for D, instead their difference $\Delta Q=Q_D-Q_L$ becomes positive and is enlarged as $r$ increases within this region. This means that players hesitate in the choice of action D or L, leading to increasing uncertainty of the environment.
This explains the non-monotonic trend observed in Fig.~\ref{fig:NonMonotonic}(b), where the power of long-term vision doesn't work anymore and players prioritize their immediate rewards.

Outside of this non-monotonic region, players no longer get vacillated between L and D as the environmental information becomes certain, and the expectation factor for the future again plays its role in the emergence of cooperation. By further increasing in $r$, the advantage of D over L becomes more obvious, and the same is also true for $Q_C>Q_D$, leading to the decreasing fraction of L and increasing trend in C.

 \section{5. Conclusion and discussion} \label{sec:discussion}

In summary, we investigate the original PGG and PGG with voluntary participation (VPGG) within the paradigm of reinforcement learning, where each player acts following a Q-learning algorithm.
Different from most previous works where the self-regarding Q-learning algorithm is employed, we incorporate environmental information into the state by comparing the numbers of cooperators and defectors in their neighborhoods.
We observe that the thresholds for cooperation emergence in both cases occurs are much reduced compared to the case in the IL framework. The cooperation emergence is found to be most likely to occur in VPGG. Unexpectedly, a non-monotonic dependence of cooperation prevalence on the gain factor is revealed in the VPGG.
By the analysis of the evolution of state, action preference, rewards, and the evolution of the Q-table, we clarify the mechanism of cooperation in both cases and the reason behind the non-monotonic dependence in VPGG.

Our research reveals significant differences in the mechanisms of cooperation emergence between the traditional IL and our RL. 
Within RL, the emergence of cooperation does not rely on the cooperator clustering, the so-called network reciprocity~\cite{Nowak1992Evolutionary}, instead by incorporating the historical experiences, present rewards, and the expected payoffs in the future, players avoid complete defection and cooperators emerge in the form of scattered distribution.
Furthermore, in the VPGG, the choice of the loner (non-participating players) for the individuals significantly inhibits the survival of defectors, and players turn to cooperators when the loners are dominating. 

Unexpectedly, there is a non-monotonic interval where the number of cooperators decreases with the gain factors $r$. Our analysis reveals that this is due to the enhanced environmental uncertainty, reducing the expectation of future rewards, and cooperation is thus degraded as players put more weight on the present rewards. This uncertainty induced risk aversion disappears as the gain factor $r$ is further increased where the environment becomes less uncertain, and the cooperation prevalence continues to increase with increasing $r$.

Till now, reinforcement learning has been employed to explain various socio-economic activities beyond cooperation, including trust~\cite{Zheng2024decoding}, resource allocation~\cite{Andrecut2001q,Zhang2019reinforcement}, and other human behaviors~\cite{Zhang2020Oscillatory,Tomov2021multi,Shi2022analysis}. In fact, there is abundant experimental evidence in neuroscience~\cite{Daeyeol2012Neural,Antonio2008A} indicating that many decision-making processes of humans fall into RL paradigm. Notice that, we do not devalue social learning~\cite{Olsson2020neural}. We conjecture that, in our daily lives both paradigms are relevant, and each works in some specific scenarios and together they underpin most of our decision-makings. Of course, this speculation requires further validation through behavioral experiments. We believe that the research conducted in this work is crucial for elucidating some basic questions along this research line.

\section{Acknowledgments}
This work was supported by the National Natural Science Foundation of China [Grants Nos. 12075144,12165014], and Fundamental Research Funds for the Central Universities (GK202401002). 
ZGZ is supported by Excellent Graduate Training Program of Shaanxi Normal University [Grants No. LHRCTS23064].

\appendix

\section{Appendix A: Phase diagrams of the learning parameters}\label{sec:appendixA}

To systematically examine the impact of the learning parameters on the cooperation evolution, we provide phase diagrams for both PGG and VPGG with Q-learning algorithm.

\begin{figure}[bpth]
\centering
\includegraphics[width=0.8\linewidth]{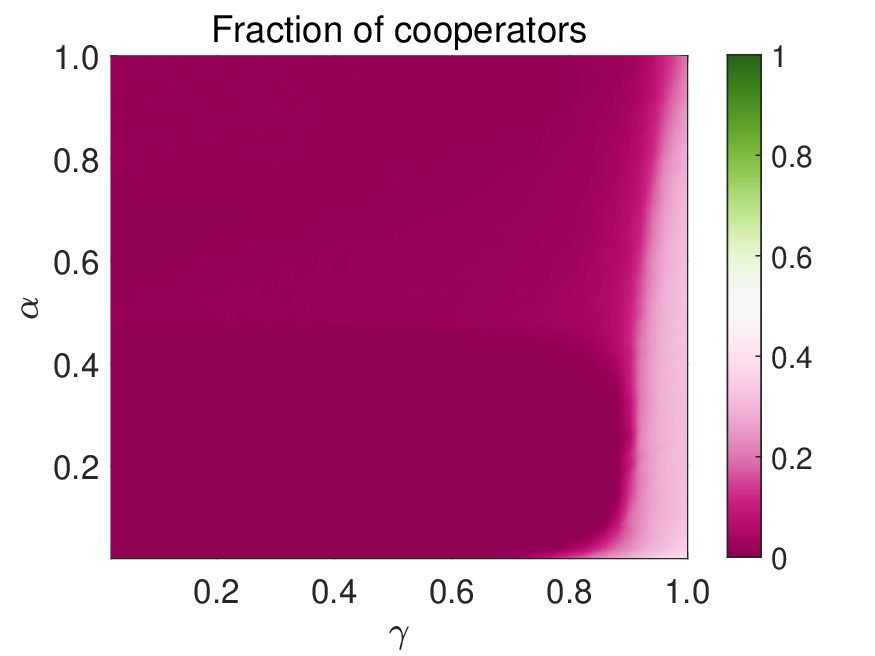}
\caption{(Color online)  The color-coded stationary fraction of cooperators in the domain ($\gamma$, $\alpha$) for PGG with the Q-learning algorithm. 
Cooperation emerges at the corner of large $\gamma$ and small $\alpha$, indicating that when individuals focus on both historical experience and the long-term vision, cooperation emerges.  Each data is averaged 100 times after a transient of $5\times10^5$. 
Other parameters: $\epsilon=0.01$, $r=3.32$, and the population size $N=100\times100$.
}
\label{fig:diagramINFPGG}
\end{figure}

In PGG with Q-learning, we find a prominent emergence of cooperation is observed in a physically meaningful region, see the phase diagram Fig.~\ref{fig:diagramINFPGG}, it reports the final fractions of cooperators in the learning parameter domain ($\gamma$, $\alpha$). As shown, there is a white region where the cooperation emerges, where the learning rate $\alpha$ is small and the discount factor $\gamma$ is large. This observation means that when individuals focus on both historical experience and long-term vision, cooperation is likely to emerge.  Otherwise, either a forgetful property (a large $\alpha$) and/or a short-term vision (a small $\gamma$) lead to the failure of its emergence.

\begin{figure}[htbp!]
\centering
\includegraphics[width=0.8\linewidth]{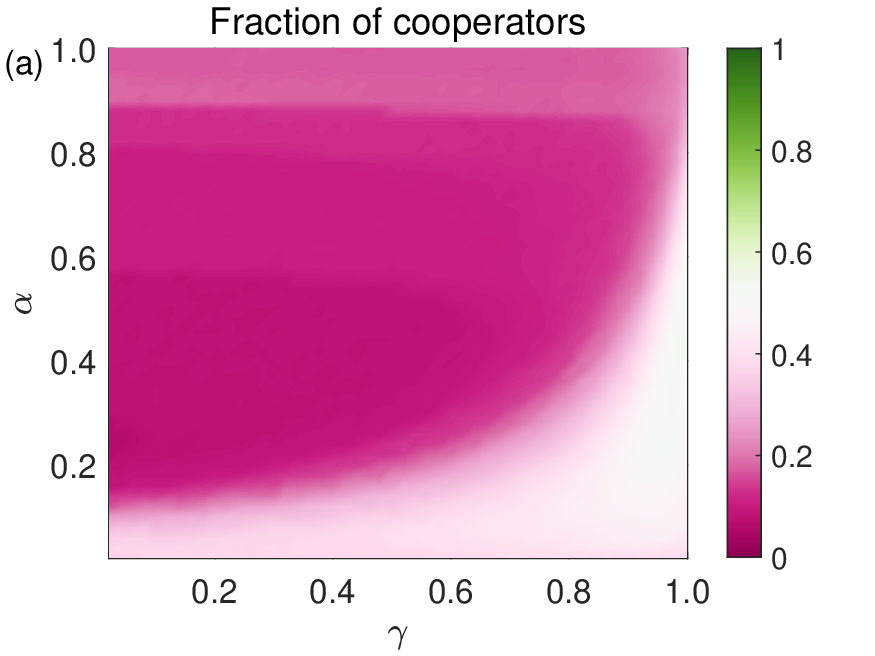}
\includegraphics[width=0.8\linewidth]{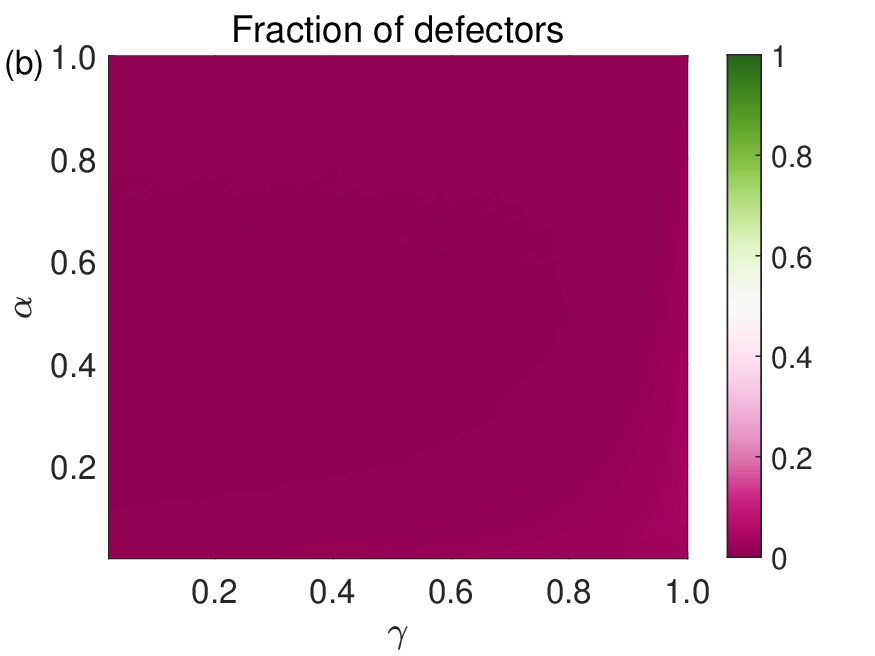}
\includegraphics[width=0.8\linewidth]{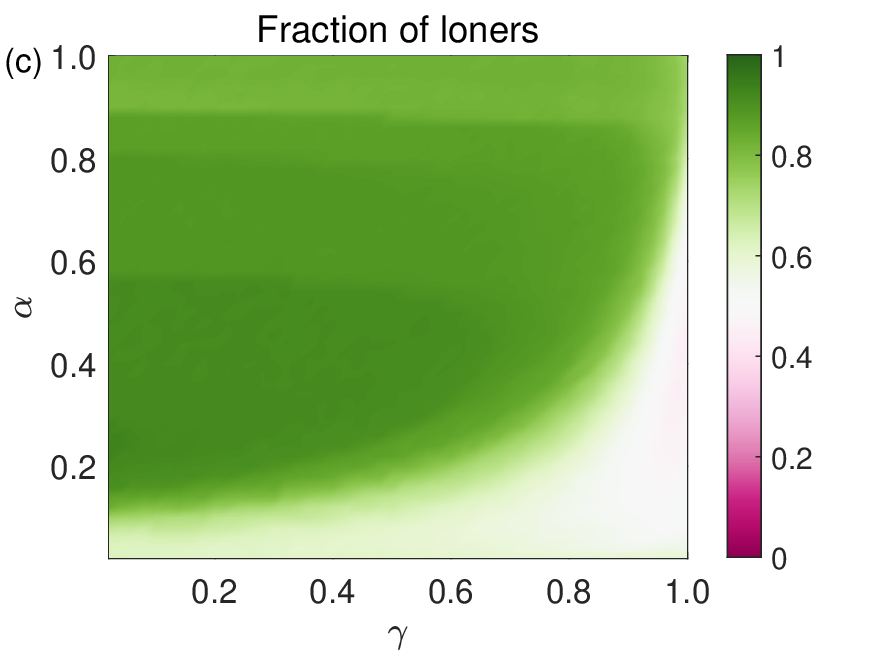}
\caption{(Color online) The color-coded stationary fraction of the three strategies in the domain ($\gamma$, $\alpha$) for VPGG with the Q-learning algorithm. 
Cooperation emerges at the corner of large $\gamma$ and small $\alpha$, showing that when individuals focus on both historical experience and the long-term vision, cooperation emerges.  
Each data is averaged 100 times after a transient of $5\times10^5$. 
Other parameters: $\epsilon=0.01$, $r=2.11$, and the population size $N=100\times100$.
} 
\label{fig:diagramInFVPGG}
\end{figure}

Similar results can also be observed in the VPGG, as shown in Fig.~\ref{fig:diagramInFVPGG}. The emergence of cooperators fails with the decrease in $\gamma$ and/or the increase in $\alpha$. However, the fraction of defectors remains small across the entire parameter space. This result is easily understandable because, with the introduction of the loner strategy, individuals tend to choose strategy L with a higher payoff rather than the all-defect strategy with zero payoffs as we discuss above. 

These observations are in line with our previous findings revealed in the emergence of cooperation in the prisoners' dilemma~\cite{Ding2023Emergence} or the donation game~\cite{Zhang2024emergence}, and the emergence of trust~\cite{Zheng2024decoding}.


\section{Appendix B: Payoff analysis in VPGG}\label{sec:appendixB}

For a more intuitive understanding of the transition in individual action preferences from the perspective of payoffs, we provide payoffs for different strategies in three states $s_{1,2,3}$, as shown in table~\ref{tab:payoffA} to table~\ref{tab:payoffC}.

Table~\ref{tab:payoffA} shows payoffs for individuals in state $s_1$ where $n_C>n_D$. Only in scenario $(nC, 0D)$ does choosing cooperation yields a higher payoff than choosing the loner. This implies that in an environment where defectors exist, cooperators are prone to shifting toward being loners.

Table~\ref{tab:payoffB} displays payoffs for individuals in state $s_2$ where $n_C=n_D$. Choosing L consistently yields the highest payoff, except in the scenario $(0C, 0D)$, where all individuals around are loners, facilitating the emergence of cooperation.

Table~\ref{tab:payoffC} presents payoffs for individuals in state $s_3$ where $n_C<n_D$. Choosing L consistently yields the highest payoff, suggesting individuals typically opt for higher payoff strategies when in this state.

\begin{table}[htpb!]
\centering
\begin{tabular}{cccc}
\arrayrulecolor{tabcolor2}\toprule [1.4pt]
\hline
\diagbox{$s_1$}{Action}& C ($a_{1}$) & D ($a_{2}$) & L ($a_{3}$) \\
\midrule [0.5pt]
\hline
1C, 0D & $1.11$ & $1.055$ & $1$ \\
\rowcolor{gray!40}2C, 0D & $1.11$ & $1.406$ & $1$ \\
2C, 1D & $0.5825$ & $1.055$ & $1$\\
\rowcolor{gray!40}3C, 0D & $1.11$ & $1.5825$ & $1$ \\
3C, 1D & $0.688$ & $1.266$ & $1$\\
\rowcolor{gray!40}4C, 0D & $1.11$ & $1.688$ & $1$ \\
\hline
\bottomrule[1.4pt]
\end{tabular}
\caption{In the VPGG, the average payoffs when choosing different actions in state $s_1$, where 
$n_C > n_D$. Parameter: $r=2.11$.}
\label{tab:payoffA}
\end{table}

\begin{table}[htpb]
\centering
\begin{tabular}{cccc}
\arrayrulecolor{tabcolor2}\toprule [1.4pt]
\hline
\diagbox{$s_2$}{Action}& C ($a_{1}$) & D ($a_{2}$) & L ($a_{3}$) \\
\midrule [0.5pt]
\hline
0C, 0D & $1.11$ & $0$ & $1$ \\
\rowcolor{gray!40}1C, 1D & $0.406$ & $0.703$ & $1$ \\
2C, 2D & $0.266$ & $0.844$ & $1$\\
\hline
\bottomrule[1.4pt]
\end{tabular}
\caption{In the VPGG, the average payoffs when choosing different actions in state $s_2$, where $n_C=n_D$. Parameter: $r=2.11$.}
\label{tab:payoffB}
\end{table}

\begin{table}[htpb]
\centering
\begin{tabular}{cccc}
\arrayrulecolor{tabcolor2}\toprule [1.4pt]
\hline
\diagbox{$s_3$}{Action}& C ($a_{1}$) & D ($a_{2}$) & L ($a_{3}$) \\
\midrule [0.5pt]
\hline
0C, 4D & $-0.578$ & $0$ & $1$ \\
\rowcolor{gray!40}0C, 3D & $-0.4725$ & $0$ & $1$ \\
0C, 2D & $-0.296$ & $0$ & $1$\\
\rowcolor{gray!40}0C, 1D & $0.055$ & $0$ & $1$ \\
1C, 2D & $0.055$ & $0.5275$ & $1$\\
\rowcolor{gray!40}1C, 3D & $-0.156$ & $0.422$ & $1$ \\
\hline
\bottomrule[1.4pt]
\end{tabular}
\caption{In the VPGG, the average payoffs when choosing different actions in state $s_3$, where $n_C<n_D$. Parameter: $r=2.11$.}
\label{tab:payoffC}
\end{table}

\section{Appendix C: Dependence on $\gamma$ outside of the non-monotonic interval}\label{sec:appendixC}

\begin{figure}[bpth]
\centering
\includegraphics[width=0.8\linewidth]{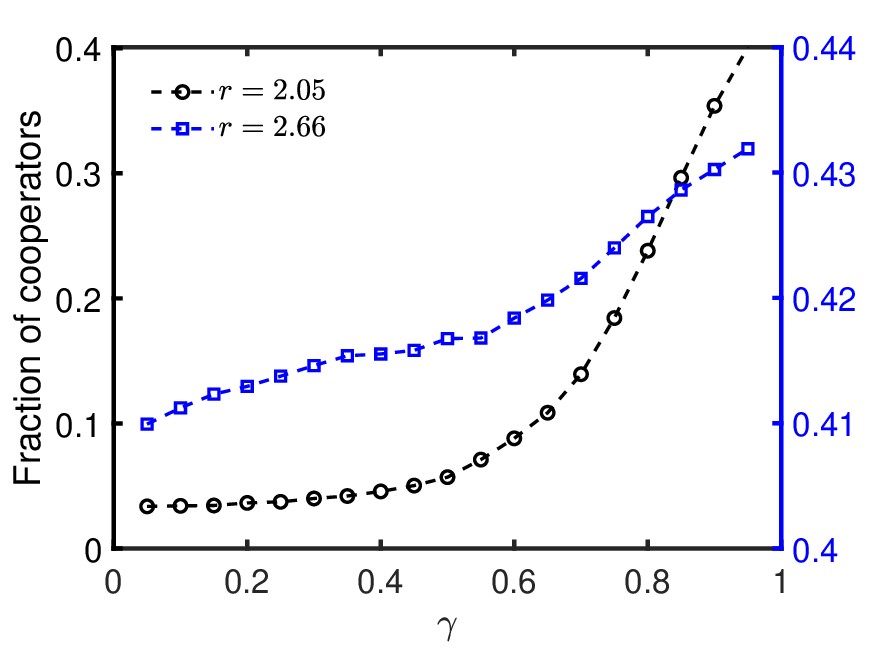}
\caption{(Color online)  The fraction of cooperators outside of the non-monotonic interval. We can see that the fraction of cooperation increases with the increase of $\gamma$.
Other parameters: $\epsilon=0.01$, $\alpha=0.1$, and the population size $N=100\times100$.
}
\label{fig:Deviate_nonmonotonic}
\end{figure}

As a comparison, here we present the fraction of cooperators outside of the non-monotonic interval (at $r=2.05$ and $r=2.66$) as the discount factor $\gamma$ varies, as shown in Fig.~\ref{fig:Deviate_nonmonotonic}. We can observe that the fraction of cooperators continues to increase with the increasing $\gamma$, quantitatively different from the observations made in Fig.~\ref{fig:NonMonotonic}(b) in the non-monotonic interval. This indicates that individuals valuing future rewards are conducive to the emergence of cooperation, which aligns with our previous findings~\cite{Ding2023Emergence, Zheng2024decoding}, where a higher expectation (i.e. a larger $\gamma$) always leads to a better outcome of evolution.

\bibliography{PGG}
\end{document}